\documentclass[twocolumn,showpacs,preprintnumbers]{revtex4}
\usepackage{graphicx}
\usepackage{epsf}
\usepackage{dcolumn}
\usepackage{bm}

\begin{document}

\preprint{APS/123-QED}
\title{Decoupling of orbital and spin degrees of freedom in Li$_{1-x}$Na$_{x}$NiO$_{2}$\\}

\author{M. Holzapfel$^{1}$$^{*}$, S. de Brion$^{1}$, C. Darie$^{2}$, P. Bordet$^{2}$, E. Chappel$^{1}$, G. Chouteau$^{1}$, P. Strobel$^{2}$, A.\ Sulpice$^{3}$ and M. D. N\'{u}\~{n}ez-Regueiro$^{4}$}
\affiliation{$^{1}$Grenoble High Field Magnetic Laboratory, CNRS and MPI-FKF, BP 166, 38042 Grenoble, France\\
$^{2}$Laboratoire de Cristallographie, CNRS, BP 166, 38042 Grenoble, France\\
$^{3}$Centre de Recherche sur les Tr\`{e}s Basses Temp\'{e}ratures, CNRS, BP 166, 38042 Grenoble, France\\
$^{4}$Laboratoire de Physique des Solides, B\^{a}timent 510, Universit\'{e} Paris-Sud, 91405 Orsay, France\\
$^{*}$present address: Paul-Scherrer-Institut CH-5232 Villigen
PSI, Switzerland\\}
\date{\today}

\begin{abstract}
In the Li$_{1-x}$Na$_{x}$NiO$_{2}$ solid solutions three different single
phase regions exist: for $x\geq0.9$, for $x\approx0.7$ and for $x\leq0.3$.
 Although the intermediate compound does not show the cooperative
Jahn-Teller transition of NaNiO$_{2}$, its magnetic properties remain very
similar with, in particular, the low temperature 3D magnetic ordering.
Therefore, the strong coupling between orbital and spin degrees of freedom,
characteristic of other oxides like perovskites, and usually invoked to
explain the absence of both long-range orbital and magnetic ordering in LiNiO$_{2}$,
does not to take place in these layered compounds with $\sim $90$^{\circ }$
super-exchange bonds. We also discuss the relevance of the O crystal field
splitting induced by the trigonal distortion, in generating AFM in-plane Ni-Ni
 interactions.
\end{abstract}

\maketitle



The interplay between orbital and spin degrees of freedom in transition
metal (TM) oxides can yield peculiar magnetic structures and so is a
subject under active research \cite{tokura}. In this context, the absence of
both long-range magnetic and orbital ordering in layered LiNiO$_{2}$, indeed clearly
observed in isomorphic and isoelectronic NaNiO$_{2}$ \cite{bongers,chappel,chappelbis},
is especially puzzling. Recently there have been attempts to explain these
curious different behaviours, but the situation is still unsettled. Most models
\cite{kitaoka,li,feiner,mila,vernay} assume an important coupling between the
spin and orbital states, considering that frustration or the particular
orbital ordering in the triangular lattice of LiNiO$_{2}$ inhibits the
stabilization of a 3D magnetic ordering. Here, our
synthesis and study of intermediate Li$_{1-x}$Na$_{x}$NiO$_{2}$ compounds
shows that the macroscopic Jahn-Teller (JT) transition observed in
NaNiO$_{2}$ can be killed by a small amount of Li replacing Na ions.
Paradoxically, the magnetic properties exhibited by Li doped NaNiO$_{2}$
remain very close to those displayed by the pure NaNiO$_{2}$.
Our experimental observations support previous analysis \cite{nunez}
and recent theoretical calculations \cite{mostovoy} showing that,
in the $\sim $90$^{\circ }$ TM-O-TM systems, orbitals and spins are essentially
decoupled. Departure from the ideal 90$^{\circ }$ angle, contribution from
extra overlapings \cite{vernay} or crystal field splitting effects
\cite{dare} have been invoked to strengthen the coupling between spins and orbitals in
these systems, but the results reported here indicate that those hypothesis
are not appropriate in the case of LiNiO$_{2}$.\newline

The $\alpha$-NaFeO$_{2}$ structure of LiNiO$_{2}$ has planes of Ni
magnetic ions arranged in a triangular network. No long range magnetic
ordering has been reported for this compound. The Ni$^{3+}$ ions are in
the low spin configuration (t$_{2g}^{6}$e$_{g}^{1}$) with s=1/2 on the
doubly degenerate $e_{g}$ level. The Goodenough-Kanamori-Anderson
 rules yield ferromagnetic (FM) spin exchange couplings between
nearest-neighbour Ni ions in the same plane. The measured Curie-Weiss
temperature indicates the predominance of FM interactions.
Interestingly, it has been recently pointed out \cite{dare} that
the trigonal crystal field splitting of the O$2p$ orbitals could induce also
antiferromagnetic (AFM) exchange integrals in the Ni planes.
That would give a microscopic foundation for the coexistence of
intra-layer FM and AFM couplings in LiNiO$_{2}$, as considered by some
authors \cite{reynaud,vernay}. Concerning the orbital occupation,
no  JT transition is observed.\\
On the other hand, NaNiO$_{2}$ undergoes a cooperative JT
ordering of the Ni$^{3+}$ ions at 480K reducing its symmetry from
rhombohedral to monoclinic. It presents a long range AFM order below
20K \cite{bongers,chappel}. It has been then speculated that a different
oxygen crystal field splitting would inhibit these AFM in plane Ni-Ni
interactions in NaNiO$_{2}$, yielding its different magnetic properties \cite{dare}.\\

We perform here a crystallographic and magnetic study of
LiNiO$_{2}$, NaNiO$_{2}$ and the new Li$_{0.3}$Na$_{0.7}$NiO$_{2}$ \cite{matsumura}.
Our main point is that samples with this intermediate composition do not undergo the
JT transition of NaNiO$_{2}$, but keep a rhombohedral structure like LiNiO$_{2}$ even
at low temperature. In spite of this fact, they achieve the long range magnetic ordering
of NaNiO$_{2}$. This shows the decoupling between orbital and spins for these
oxides with $\sim $90$^{\circ }$ bonds. Furthermore, the quantification of the trigonal
distortion in these 3 compounds also allow us to discuss its relevance in generating
in-plane AFM interactions in LiNiO$_{2}$, conjectured to suppress the magnetic
ordering \cite{dare}.\\

\begin{figure}[t]
\setlength{\epsfysize}{4cm} \centerline{\epsffile{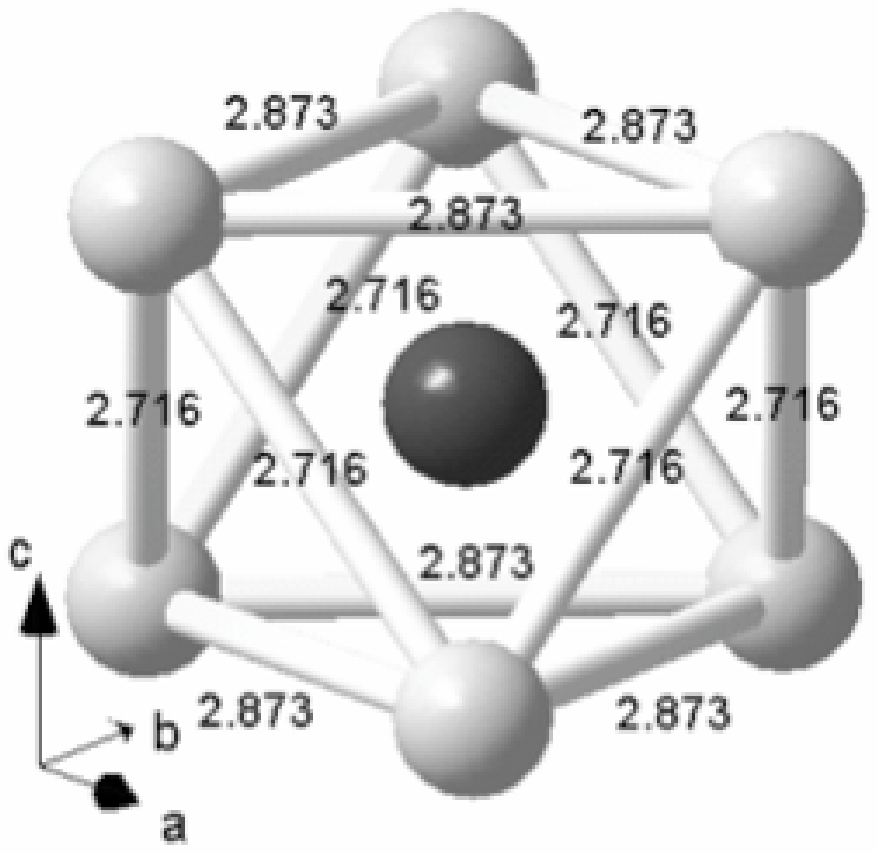}}
\caption{the trigonal distorted NiO$_{6}$ octahedra in LiNiO$_{2}$
with O-O distances.}
\end{figure}

\begin{figure}[b]
\setlength{\epsfysize}{7cm} \centerline{\epsffile{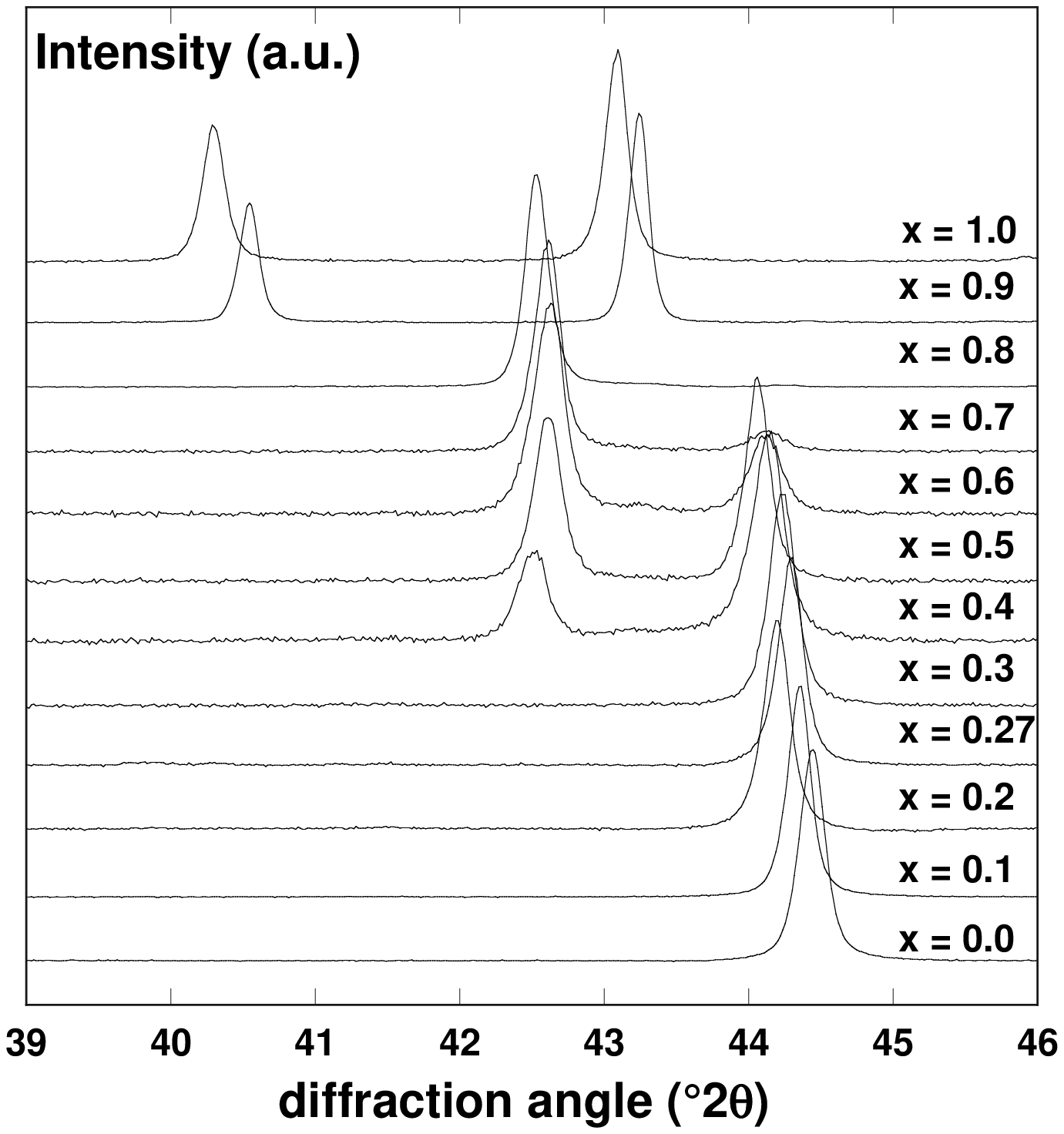}}
\caption{x-ray diffractograms as a function of the nominal composition Li$%
_{1-x}$Na$_{x}$NiO$_{2}$.}
\end{figure}

The products were obtained by classical high-temperature solid-state
reaction. They were synthesized from Na$_{2}$O$_{2}$, NiO and LiOH$\cdot$H$_{2}$O powders.
The starting materials were mixed in an argon atmosphere with a 10\% excess
of (Na$+$Li) to account for volatilization losses. The mixture was heated
under flowing oxygen at 680$^{\circ}$C for 24h. Twelve samples with compositions
(Li$_{1-x}$Na$_{x}$)NiO$_{2}$ with 0$\leq x \leq$1 ($x$ denotes the nominal Na content)
were prepared. Their cationic compositions were analyzed by atomic absorption
spectroscopy. The crystal structure was studied by x-ray powder diffraction (XPD).
In order to detect possible magnetic ordering, neutron powder diffraction
(NPD) data were collected down to 1.5K on the CRG-D1B instrument of the ILL.
The x-ray and neutron diffractograms were refined by Rietveld technique using
the Fullprof software \cite{rodriguez}.\\

It is not possible to synthesize single phase samples for arbitrary Li/Na
ratio. As shown in Fig.2 and reported by Matsumura et al.\cite{matsumura},
the LiNiO$_{2}$/NaNiO$_{2}$ phase diagram contains 3
different single phase solid solution regions, in between which, two phase
mixtures are observed. For $x\leq $0.3, the R$\bar{3}$m $\alpha $-NaFeO$_{2}$
structure type of LiNiO$_{2}$ is found (RII phase), where Li, Ni and O atoms
occupy the 3a (0 0 0), 3b (0 0 1/2) and 6c (0 0 z$\approx $0.24) positions,
respectively. For $x\geq $0.9, the monoclinic C2/m phase of NaNiO$_{2}$ is
observed, with substitution of Li$^{+}$ for Na$^{+}$ ions. In this phase,
the Na, Ni and O atoms occupy the 2d (0 1/2 1/2), 2a (0 0 0) and 4i (x$%
\approx $0.28 0 z$\approx $0.8) positions. In a small concentration range about $x\approx
$0.8, a new rhombohedral phase (RI phase) is obtained, which has the same
structural arrangement as LiNiO$_{2}$, but quite different cell parameters:
the c/a ratio is $\approx $5.24 instead of $\approx $4.94 for LiNiO$_{2}$.
The exact cationic composition of this phase  found both by Rietveld
refinement of the x-ray data and atomic absorption is Li$_{0.30(1)}$Na$_{0.70(1)}$.
No structural phase transition could be detected by NPD for (Li$_{0.3}$Na$_{0.7}
$)NiO$_{2}$ down to 1.5K. Therefore, this phase remains rhombohedral in this
temperature range and does not undergo a cooperative JT ordering. In this
respect, it behaves like LiNiO$_{2}$ and not like NaNiO$_{2}$.\\

\begin{table}[t]
\caption{room temperature crystallographic and low temperature magnetic
parameters for the three phases.}
\label{tab:table1}%
\begin{ruledtabular}
\begin{tabular}{ccccc}
 &LiNiO$_{2}$& Li$_{0.3}$Na$_{0.7}$NiO$_{2}$& NaNiO$_{2}$&\\
\hline
space group  & R$\bar{3}$m & R$\bar{3}$m  & C2/m\\
cell param.  & a=2.8727(3) & a=2.9410(1)  & a=5.3208(3)\\
             & c=14.184(2) & c=15.4082(1) & b=2.8440(2)\\
             &             &              & c=5.5818(4)\\
             &             &              & $\beta$=110.49(1)\\
\hline
Na content  & 0.009(3)  & 0.698(4)     & 1.0\\
\hline
positional   & x(O)=0.24248(8) & x(O)=0.2324(2) & x(O)=0.282(2)\\
parameters   &                 &                & z(O)=0.799(2)\\
\hline
R$_{wp}$,R$_{Bragg}$& 3.91,7.28 &3.24,6.6       & 3.78,8.11\\
\hline
Ni-O (\AA)   & 6x1.977         & 6x1.977        & 4x1.93, 2x2.17\\
\hline
O-O (\AA)    & 6x2.716         & 6x2.644        & 4x3.02, 2x2.84\\
             & 6x2.873         & 6x2.941        & 4x2.78, 2x2.60\\
\hline
O-Ni-O ($^{\circ}$) & 6x93.2   & 6x96.1         & 4x94.8, 2x95.1\\
                    & 6x86.8   & 6x83.9         & 4x85.2, 2x84.9\\

\hline
\hline
$T_{CW}$            & +26K     & +40K           & +36K\\
\hline
$T_{N}$ or T$_{SG}$ &   9K     &  25K           &  20K\\
\hline
$H_{C0}$ at 4K      &          & 0.05T          &  1.8T\\
\hline
$H_{C1}$            &          &    5T          & 7T\\
\hline
$H_{sat}$           & $>$23T   &   19T          & 13T\\
\hline
$H_{E}$             &          &  9.5T          & 6.5T\\
\hline
$H_{A}$             &          &0.2mT           & 250mT\\
\end{tabular}
\end{ruledtabular}
\end{table}

To better understand  the magnetism of these compounds, it is interesting to
compare the structural arrangements of the NiO$_{2}$ octahedral layers (Table I).
The R$\bar{3}$m LiNiO$_{2}$ structure can be described by starting from the
F-centered cubic NiO structure and replacing every other Ni plane
perpendicular to one of the 3-fold axes by a Li plane, forming NiO$_{2}$
edge-shared octahedral layers separated by Li$^{+}$ cations planes. This is
accompanied by a trigonal distortion of the NiO$_{6}$ octahedra brought
about by the difference of charge and size between Li$^{+}$ and Ni$^{3+}$
cations. These octahedra are compressed along the c axis as shown in Fig.1.
The six Ni-O distances remain equal ($\approx $1.98\AA ), but the six O-O
distances linking oxygen anions from the plane above and below the Ni cation,
called d(O-O)$_{c}$, become shorter ($\approx $2.72\AA) than those in
the ab plane, d(O-O)$_{ab}$ ($\approx $2.87\AA ). At the same time, the
O-Ni-O angles between oxygen anions from the same plane above or below the
Ni cation opens to 93.2$^{\circ }$, while the ones between oxygen anions
from the two planes close down to 86.8$^{\circ }$. This distortion can be
characterized by the ratio $\gamma=$d(O-O)$_{c}$/d(O-O)$_{ab}$, and is
0.945 for LiNiO$_{2}$.\\
Although the structural arrangement is the same for Li$_{0.3}$Na$_{0.7}$NiO$_{2}$,
the interatomic distances and angles are markedly modified because
of the steric effect due to the large size difference between Na$^{+}$ and Li$^{+}$
cations (ionic radius=1.02\AA  and 0.76\AA). Due to charge balance,
the 6 Ni-O distances are almost unchanged (1.98\AA), but the 2 O-O
distances and angles defined above become $\approx $2.94\AA , 2.64\AA ,
96.1$^{\circ }$ and 83.9$^{\circ }$, respectively. $\gamma=$0.899,
indicates a much stronger trigonal distortion than for LiNiO$_{2}$.

Very similar Ni-O and O-O distances and O-Ni-O angles can be calculated in
the isostructural high temperature form of NaNiO$_{2}$ from the NPD data of
Chappel et al. \cite{chappelbis} (at 565K: Ni-O : 6x1.98\AA; O-O : 6x2.96\AA,
6x2.63\AA,  $\gamma=$0.899 ; O-Ni-O : 6x96.8$^{\circ}$, 6x83.2$^{\circ}$).
At room temperature, in the monoclinic phase, the NiO$_{6}$ octahedra
become JT distorted: 4 Ni-O distances become shorter (1.93\AA) and 2 longer
(2.17\AA). The average O-O distances become d(O-O)$_{c}\simeq$2.72(0.1)\AA,
 and d(O-O)$_{ab}\simeq$2.96(0.1)\AA, and the average O-Ni-O angles
$\simeq$95$^{\circ}$ and $\simeq$85$^{\circ}$. The same ratio can still be used to evaluate
the trigonal distortion: $\gamma=$0.919. By comparing these values with those
of (Li$_{0.3}$Na$_{0.7}$)NiO$_{2}$, one can conclude that in monoclinic NaNiO$_{2}$,
the JT distortion is superimposed on the trigonal one without markedly changing
the amplitude of the latter.

\begin{figure}[b]
\setlength{\epsfysize}{6cm} \centerline{\epsffile{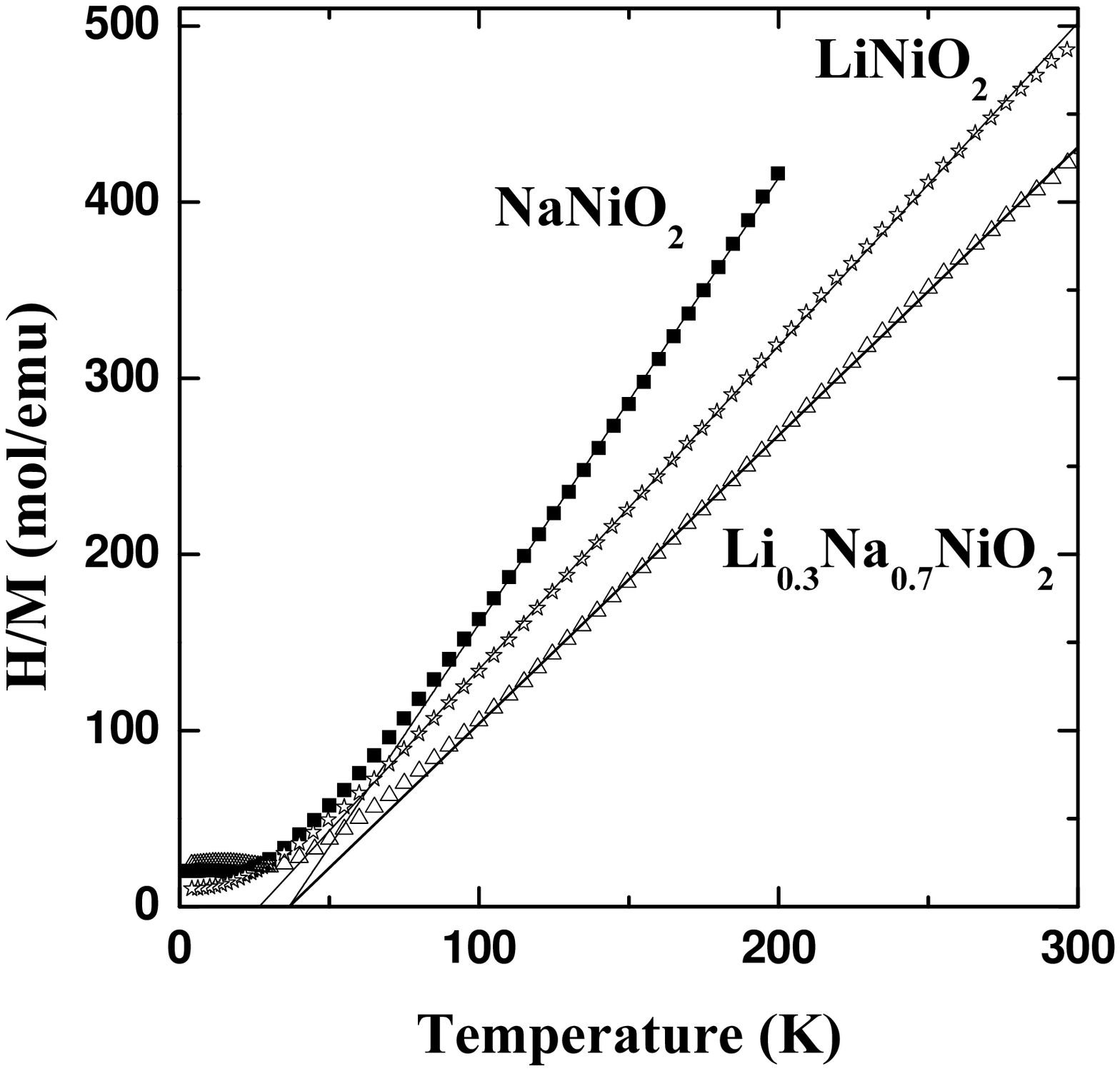}}
\caption{inverse of the susceptibility vs temperature, showing the
Curie-Weiss behaviour.}
\end{figure}

\begin{figure}[tb]
\setlength{\epsfysize}{11cm} \centerline{\epsffile{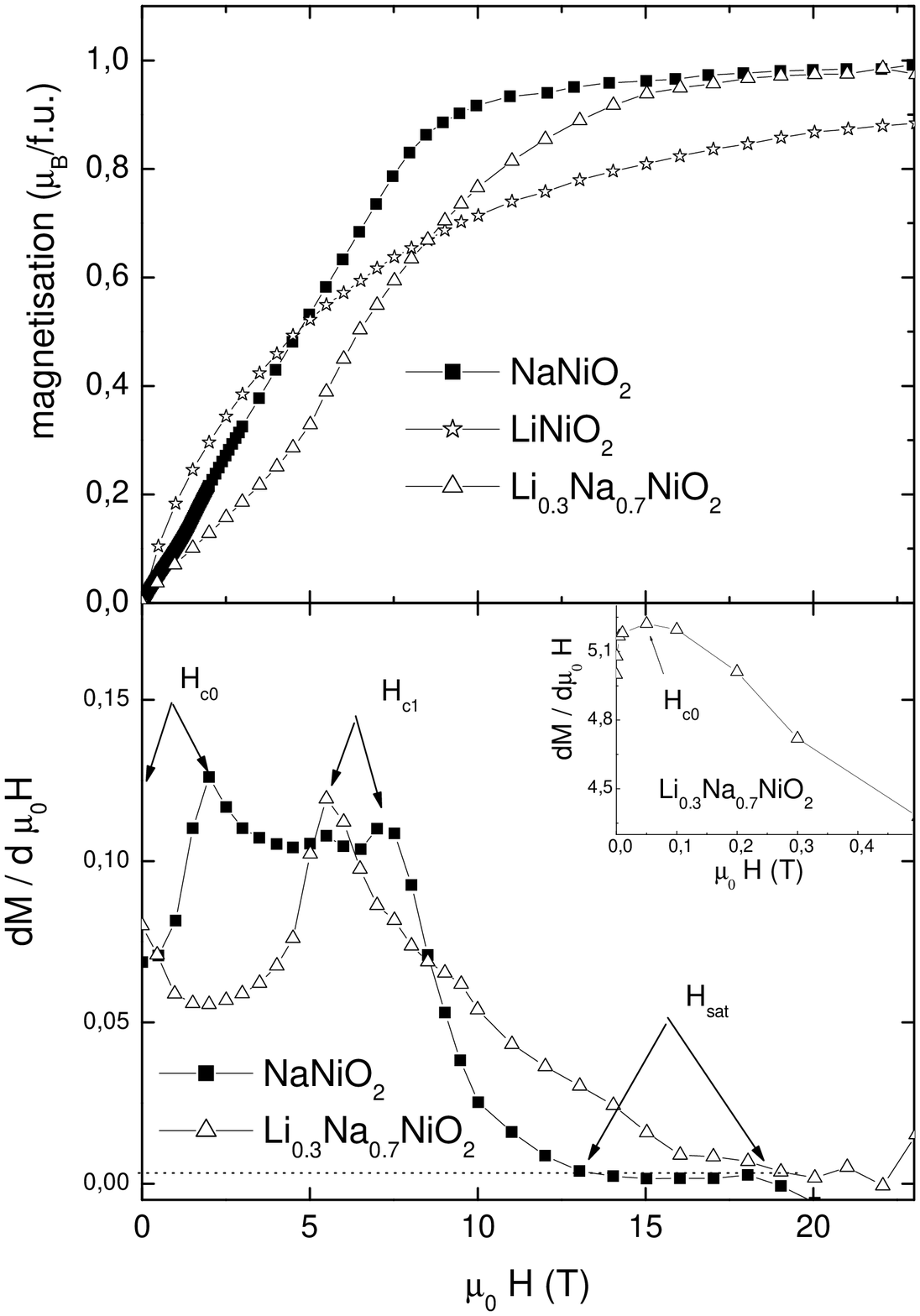}}
\vspace{0.2cm}
 \caption{field dependence (top) and field
derivative of the magnetization (bottom) at 4K, showing the 3
characteristics fields. Inset: blow up of the low field results
for Li$_{0.3}$Na$_{0.7}$NiO$_{2}$.} \label{fig:epsart}
\end{figure}

Having in mind that NaNiO$_{2}$ undergoes a collective JT transition, while
LiNiO$_{2}$ as well as Li$_{0.3}$Na$_{0.7}$NiO$_{2}$ do not, we perform now
a comparative study of their magnetic properties. The positive Curie-Weiss
temperature for all of them   (Fig.3 and Table I)  reflects the predominance
of FM interactions. The effective moment in all the high temperature orbitally disordered
phases is nearly the same ($\simeq$2.1$\mu _{B}$) while it is smaller in the
orbitally ordered phase of NaNiO$_{2}$ (1.85$\mu _{B}$).
Deviation from the free electron value (1.75$\mu _{B}$) is enhanced in the
orbital ordered phase reflecting the importance of the orbital contribution
to the Ni$^{3+}$ magnetic moment.\\
The 3 compounds show a maximum
in the susceptibility at low temperature: 20K for NaNiO$_{2}$, 25K for
Li$_{0.3}$Na$_{0.7}$NiO$_{2}$ and 9K for LiNiO$_{2}$. In fact,
 LiNiO$_{2}$ is never stoichiometric: the exact formula being
Li$_{1-\epsilon }$Ni$_{1+\epsilon }$O$_{2}$ wih extra Ni ions in the Li plane.
This gives rise to frustrated magnetic interactions between the Ni planes
and prevents any long range magnetic order yielding instead spin glass
behaviour \cite{chappelbisbis}. In NaNiO$_{2}$ and Li$_{0.3}$Na$_{0.7}$NiO$_{2}$
which are stoichiometric (no Ni ions are present in the Na/Li planes),
the field dependence of the magnetization at 4K indicates long range magnetic order
for both compositions (Figs.4a and b).
NaNiO$_{2}$ is considered as an A type antiferromagnet with FM Ni layers
coupled AF below T$_{N}\simeq $20K \cite{bongers,chappel,chappelbis}.
However our NPD measurements failed to detect such a simple magnetic structure,
in spite of the adequate sensitivity of the D1B spectrometer.
A careful analysis of the magnetization data reveals indeed a more complex
ordering: when taking its derivative as a function of the applied magnetic
field (Fig.4b), clearly 3 characteristic fields are present: $H_{c0}$,
$H_{c1}$ and $H_{sat}$ instead of 2 expected for an A type antiferromagnet
($H_{c0}$, and $H_{sat}$) \cite{herpin}. The same behaviour is observed in
Li$_{0.3}$Na$_{0.7}$NiO$_{2}$, with $H_{c0}$, $H_{c1}$ lowered and $H_{sat}$
enhanced (Table I). Similarly no magnetic diffraction peak was detected by NPD.
So, we can conclude that both NaNiO$_{2}$ and Li$_{0.3}$Na$_{0.7}$NiO$_{2}$ undergo
the same long range magnetic transition, this is the relevant point for our discussion
concerning the decoupling of the orbital and spin degrees of freedom in these systems.

In a first approach, this common magnetic structure can still be described as an
AF stacking of FM planes as previously proposed \cite{bongers,chappel,chappelbis}.
Then only $H_{sat}$ and $H_{c0}$ are relevant and we can estimate
$H_{sat}\simeq 2H_{E}$ and $H_{c0}=H_{sf}\simeq \sqrt{2H_{E}H_{A}}$, where $H_{E}$
is the AF exchange field (between the Ni layers) and $H_{A }$ the
anisotropy field which aligns the magnetic moment
in a given direction \cite{herpin}. Table I yields all these measured and calculated
quantities. Note the significant decrease of $H_{A}$: while NaNiO$_{2}$ is
an easy plane antiferromagnet, Li$_{0.3}$Na$_{0.7}$NiO$_{2}$ is
more isotropic. This can be well explained by the presence or absence of orbital
order in these compounds: the magneto-crystalline anisotropy arises from
spin-orbit coupling; in NaNiO$_{2}$ orbital order insures the presence of
preferred orientations for the magnetic moment, whereas in Li$_{0.3}$Na$_{0.7}$NiO$_{2}$
the disordered orbital occupancy gives rise to isotropic
probability for the spins orientation. From $H_{E}$ we deduce the AF
interaction J$_{AF}$: -1.3K for NaNiO$_{2}$ and  -1.9K for Li$_{0.3}$Na$_{0.7}$NiO$_{2}$.
Using also the Curie-Weiss temperature, we calculate the FM
interaction J$_{F}$: +13.3K for the former, +15.2K for the latter. Both
interactions increase slightly from NaNiO$_{2}$ to Li$_{0.3}$Na$_{0.7}$NiO$_{2}$.
J$_{AF}$ arises from super-super-exchange Ni-O-O-Ni bonds across the
Ni planes, and J$_{F}$ from $\sim $90$^{\circ }$  Ni-O-Ni in-plane bonds explained
before.\\

To summarize, while Li$_{0.3}$Na$_{0.7}$NiO$_{2}$ and NaNiO$_{2}$
have very different orbital ground state, they present a similar
magnetic ground state with similar exchange energies. The orbital
contribution can only be seen in the value of the magnetic moment
and the anisotropy field associated with Ni$^{3+}$ ions. Although
we cannot conclude about the orbital occupation in the Li
containing compounds, NaNiO$_{2}$ can most probably be the spin
model for the magnetic ground state of pure LiNiO$_{2}$
\cite{chappelbisbis}. The exact magnetic structure of NaNiO$_{2}$,
most likely a modulated one derived from the A-type
antiferromagnet, remains to be determined but this is out of the
scope of this paper. In the calculation of Dar\'{e} et
al.\cite{dare} a large enough trigonal distortion can generate AFM
interactions in the Ni planes in addition to the FM interactions.
This parameter has been quantified in our structural analysis:
LiNiO$_{2}$ has the lowest trigonal distortion ($\gamma=$0.945)
compared to Li$_{0.3}$Na$_{0.7}$NiO$_{2}$ (0.899) and to
NaNiO$_{2}$ (0.89 for the high temperature orthorhombic phase,
0.92 for the monoclinic phase). Then this mechanism should lead to
a stronger AF contribution in NaNiO$_{2}$ and Li$_{0.3}$Na$_{0.7}$NiO$_{2}$
than in LiNiO$_{2}$ while the opposite occurs according to the measured
Curie-Weiss temperatures (Table I).

In conclusion, this crystallographic and magnetic study shows that the intermediate
compound Li$_{0.3}$Na$_{0.7}$NiO$_{2}$ has the same magnetic behaviour as NaNiO$_{2}$
while its orbital behaviour is different: like LiNiO$_{2}$
it does not undergo a collective Jahn-Teller transition. So, these experimental results
establish that the orbital and spin degrees of freedom are decoupled in these systems.\\

We thank A.M. Dar\'{e}, M.V. Mostovoy and D.I. Khomskii for
discussions and O. Isnard for NPD data collected on the CRG-D1B
diffractometer at ILL. The GHMFL is "laboratoire associ\'{e} $\grave{a}$ l'Universit\'{e}
J. Fourier-Grenoble".


\begin{references}

\bibitem{tokura} Y. Tokura , and N. Nagaosa, Science {\bf5465}, 462 (2000).

\bibitem{bongers} P. F. Bongers , and U. Enz, Solid State Comm. {\bf4}, 153 (1966).

\bibitem{chappel} E. Chappel , M.D.  N\'{u}\~{n}ez-Regueiro, F. Dupont, G. Chouteau, C. Darie , and A. Sulpice, Eur.\ Phys.\ J.\ B {\bf17}, 609 (2000).

\bibitem{chappelbis} E. Chappel , M.D.  N\'{u}\~{n}ez-Regueiro, G. Chouteau, O. Isnard , and C. Darie, Eur.\ Phys.\ J.\ B {\bf17}, 615 (2000).

\bibitem{kitaoka} Y. Kitaoka , T. Kobayashi, A. Koda, H. Wakabayashi, Y. Niino, H. Yamakage, S. Taguchi, K. Amaya, K. Yamaura, M. Takano, A. Hirano , and R. Kanno, J. \ Phys. Soc. Japan {\bf67}, 3703 (1998).

\bibitem{feiner} L. F. Feiner, A. M. Oles, and J. Zaanen, Phys.\ Rev. Lett. {\bf78}, 2799 (1997);
and Phys.\ Rev.\ B {\bf61}, 6257 (2000).

\bibitem{li} Y. Q. Li , M. Ma, D.N. Shi , and F.C. Zhang , Phys.\ Rev. Lett. {\bf81}, 3527 (1998).

\bibitem{mila} M. van der Bossche , F.-C. Zhang , and F. Mila , Eur.\ Phys.\ J.\ B {\bf17}, 367 (2000);
M. van der Bossche , P. Azaria, P. Lecheminant, F. Mila , Phys.
Rev. Lett. {\bf86}, 4124 (2001).

\bibitem{vernay} F. Vernay, K. Penc, P. Fazekas, F. Mila
, cond-mat/0401122

\bibitem{nunez} M.D. N\'{u}\~{n}ez-Regueiro E. Chappel , G. Chouteau1 , and C. Delmas , Eur.\ Phys.\ J.\ B {\bf16}, 37 (2000).

\bibitem{mostovoy} M. V. Mostovoy and D. I. Khomskii, Phys.\ Rev. Lett. {\bf89}, 227203 (2002).

\bibitem{dare} A.-M. Dar\'{e}, R. Hayn, and J.-L. Richard, Europhys. Lett. {\bf61}, 803 (2003).

\bibitem{reynaud} F. Reynaud , D. Mertz, F. Celestini, J.-M. Debierre, A. M. Ghorayeb, P. Simon, A. Stepanov,
J. Voiron, and C. Delmas , Phys. Rev. Lett. {\bf86}, 3638 (2001).

\bibitem{matsumura} T. Matsumura , R. Ka nno, R. Gover, Y. Kawamoto, T. Kamiyama, B.J. Mitchell, Solid State Ionics {\bf152}, 303 (2002).

\bibitem{rodriguez} J. Rodriguez-Carvajal, Physica B {\bf192}, 55 (1993).

\bibitem{chappelbisbis} E. Chappel , M. D. Nu´n˜ez-Regueiro, S. de Brion, and G. Chouteau, V. Bianchi, D. Caurant, and N.
Baffier , Phys. Rev. B {\bf66}, 132412 (2002).

\bibitem{herpin} A. Herpin, Th\'{e}orie du Magn\'{e}tisme (Presses Universitaires de France, Paris, 1968)

\end{references}
\end{document}